\documentclass[a4paper,twocolumn,english,prb,superscriptaddress]{revtex4}

\usepackage{graphicx,graphics}
\usepackage{amsmath}
\usepackage{amssymb}
\usepackage{color}
\usepackage[update,prepend]{epstopdf}

\newcommand{\be}{\begin{equation}}
\newcommand{\beqn}{\begin{eqnarray}}
\newcommand{\eeq}{\end{equation}}
\newcommand{\ee}{\end{equation}}

\begin{document}

\title{Tunable zero-energy transmission resonances in shifted graphene bilayer}

\author{A. Daboussi}
\author{L. Mandhour}
\affiliation{Laboratoire de Physique de la Mati\`ere Condens\'ee, Facult\'e des Sciences de Tunis, Universit\'e de Tunis el Manar, Campus Universitaire Tunis, El Manar, 2092 Tunis, Tunisie.}
\author{J.N. Fuchs}
\affiliation{Laboratoire de Physique Th\'eorique de la Mati\`ere Condens\'ee, CNRS UMR 7600,
Universit\'e Pierre et Marie Curie, 4 place Jussieu, F-75252 Paris, France.}
\affiliation{Laboratoire de Physique des Solides, CNRS UMR 8502, Universit\'e Paris-Sud, F-91405 Orsay, France.}
\author{ S. Jaziri$^{1,}$}
\affiliation{Laboratoire de Physique des Mat\'eriaux, Facult\'e des Sciences de Bizerte, Universit\'e de Carthage, Jarzouna, 7021 Bizerte, Tunisie.}

\begin{abstract}

A graphene bilayer is known to perfectly reflect normally incident electrons due to their chirality. This is similar to Klein tunneling, which, in a monolayer, is instead responsible for perfect transmission at normal incidence. Stacking defaults turn each parabolic band crossing of a bilayer into pairs of Dirac cones. Here we show that, surprisingly, a stacking default (or shift) in a bilayer can result in perfect {\it transmission} at normal incidence as a result of Fabry-P\'erot type resonances {\it at zero-energy}. These constructive interferences only happen for a specific orientation of the Dirac cones with respect to the incident electron and for quantized values of their separation in reciprocal space. Our results provide a way to control transmission resonances in undoped graphene bilayer structure by adjusting the layer stacking.

\end{abstract}

\date{\today}
\maketitle

\section{Introduction}

Since the first experimental study of a graphene bilayer\cite{Novoselov,McCann}, a continuously growing interest has been devoted to both fundamental properties and possible applications of stacks of few graphene layers. Two graphene layers generally stack over the common $AB$ or Bernal stacking configuration (see below). Recently, it has been reported that several types of stacking defects can occur in natural and synthetic graphene bilayer systems \cite{{Berger},{Haas1},{Haas2},{Luican},{Li},{Miller1},{Miller2}}. For instance, chemical vapor deposition (CVD) growth of two graphene layers on metals and on the carbon face of the SiC substrates reveals natural stacking faults originating from a rotational mismatch by some twist angle between the two layers as compared to the perfect $AB$-stacked bilayer \cite{Reina}. It has also been possible to produce such rotationally faulted bilayers using different methods such as folding of mechanically exfoliated graphene samples \cite{folding}, segregation on Ni films \cite{segregation} and unzipping of carbon nanotubes \cite{unzipping}. Additionally, lateral translations of one layer with respect to the other \cite{{Ohta},{Min}} or the application of homogenous mechanical strain along some direction \cite{Falko1} in bilayer graphene are also considered as stacking defects compared to the perfect $AB$ stacking.

Several recent studies address the physics of such twisted, strained or slided bilayer graphene and reveal intriguing properties such as the observation of two symmetric low-energy van Hove singularities in the density of states with positions tunable by the relative angle between layers \cite{{Luican},{Li}}, the renormalization of Fermi velocity for the twisted bilayer \cite{{Lopes},{Trambly}} and the possibility of employing strain to annihilate two among four Dirac points \cite{Falko1}. These structural deformations have in common that they produce a change in the topology of the band structure at low-energy and are all described by a constant shift (a complex number with energy dimension) in the low-energy $AB$ bilayer Hamiltonian \cite{{Lopes},{de Gail1},{de Gail2},{Shallcross},{Falko1},{Son}}.

Understanding the effects of such a shift on the graphene bilayer properties attracted immediately a lot of theoretical and experimental interest. The low-energy dispersion of such stacking faulted graphene bilayer, named hereafter shifted graphene bilayer (SGB), can be represented by two mini Dirac cones in each valley separated by a wavevector that depends on the precise stacking defect. Such a system is very different from both the $AB$ bilayer and the monolayer. As they strongly affect the electronic band structure and hence transport properties, stacking defects could therefore be used to tailor graphene based devices.

From the point of view of transport properties, one of the most counter-intuitive and unexpected phenomena predicted and observed in graphene monolayer is the Klein tunneling effect \cite{Klein}, in which normally incident electrons move across a potential barrier with certainty regardless of the height and the width of the barrier. It has been reported that, in contrast with the monolayer case, where the complete transmission takes place exactly at normal incidence \cite{{Katsnelson_monolayer},{Klein},{Tworzydlo}}, the complete transmission through a potential barrier in $AB$ graphene bilayer occurs for some finite incident angles \cite{{Katsnelson_bilayer},{Klein},{Snyman}}. These resonances corresponding to a maximum transmission probability were studied in-depth for the monolayer in Ref. \onlinecite{Fuchs} and for the $AB$-graphene bilayer in Ref. \onlinecite{Snyman}. Two different kinds of resonance have been found: (i) evanescent waves resonances and (ii) Fabry-P\'erot resonances. First, evanescent waves resonances are a characteristic property of graphene devices whose quasi-particles are chiral and occur at either zero or finite energy in monolayer at normal incidence (this is known as Klein tunneling) \cite{{Klein},{Fuchs}}. This type of resonances also happens in perfect $AB$ bilayers for quasi-particles with finite (i.e. non-normal) incidences\footnote{We here specifically refer to incidences corresponding to a wavector such that its component parallel to the interface is $k_y=\ln (\sqrt{2}k_FL_x) /L_x$ \cite {Katsnelson_bilayer}. See below for the definition of the various quantities.} but only at zero energy \cite{Katsnelson_bilayer, Snyman}. Second, Fabry-P\'erot resonances originate from constructive interferences of multiple reflected propagating waves between double interfaces. They occur only at non-zero energy in either the monolayer or the bilayer \cite{{Snyman},{Fuchs}}. They have been observed recently in ballistic and coherent suspended monolayer devices \cite{Rickhaus}.

Understanding and controlling electron transmission resonances in SGB is therefore of importance. In the present work, we will be interested mainly in the effect of a shift on the transmission resonances through a potential barrier in undoped SGB. The paper is organized as follows. In section II, we briefly review the effective two-band low energy Hamiltonian of the SGB and discuss the validity of the description that is to be used in the following sections. In section III, we present the transmission probability calculations through a zero-energy SGB for different stacking configurations corresponding to a positive, a negative or an arbitrary complex shift. Section IV suggests an experimental implication of our results. Conclusions are given in section V.

\section{Shifted bilayer Hamiltonian}

\begin{figure*}[t]
  \centering
  \begin{tabular}{ccc}
\setlength{\unitlength}{1mm}
\begin{picture}(40,50)
\put(-50,0){\includegraphics[width=0.85\textwidth]{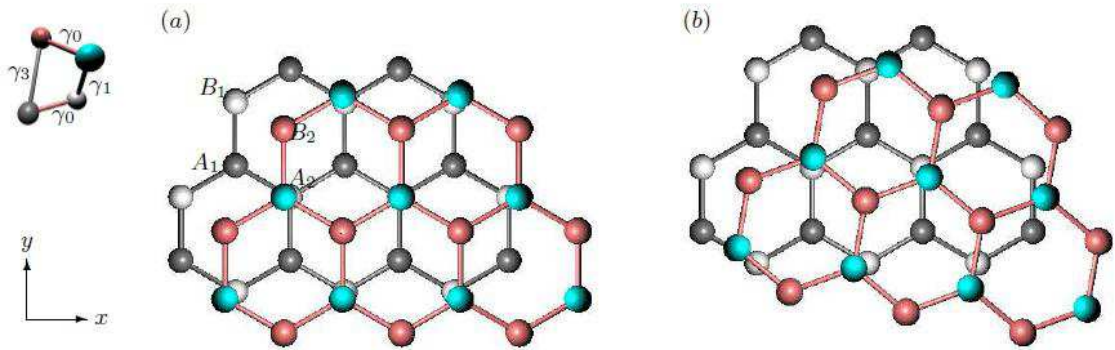}}

\end{picture}
\end{tabular}
\caption{(Color online) Top view of a perfectly $AB$-stacked (Left: (a)) and rotationnaly faulted (Right: (b)) bilayer graphene crystal. The top layers are shown in (Pink,Cyan) colors and the bottom layers are shown in (Gray,White) colors. Hopping parameters $\gamma_0$,$\gamma_1$ and $\gamma_3$ are depicted in the top left panel.}
\label{figure1}
\end{figure*}

The $AB$-bilayer graphene (BG) can be viewed as two honeycomb lattices on top of each other (see Fig.\ref{figure1}(a)).
Each layer consists of two non-equivalent sublattices with
atoms $A_{1(2)}$, $B_{1(2)}$ in the bottom (top) layer. These Bernal
stacked layers have each $A$ site surrounded by three $B$ sites,
within each layer, with intralayer hopping amplitude $\gamma_0 \approx 3 eV$.
The $A_2$ atoms are placed just above the $B_1$ ones with interlayer
coupling $\gamma_1 \approx 0.4 eV$ whereas sites $A_1$ and $B_2$ are
placed over the hexagons in the other layer and are coupled by
a hopping energy $\gamma_3 \approx 0.3 eV$. Such a stacking order gives
rise to a band structure containing two parabolic ``low-energy'' bands
touching in the Brillouin zone corners $K$ and $K'$ and two ``high-energy'' bands
split by the energy $\pm \gamma_1$. For energies lower than
$\gamma_1$, the Hamiltonian of the perfect $AB$-bilayer graphene
around the $K$ Dirac point is given by the following effective two-band Hamiltonian \cite{McCann}:\be
H_{AB}=h_{AB}+h_{TW} \ee
where the isotropic part:
 \be h_{AB}=-\frac{1}{2m_*}
\left(\begin{array}{cc} 0 & \Pi^{\dagger 2}
\\ \Pi^2 & 0 \end{array}\right) \ee
and the trigonal warping part:
\be h_{TW}= v_3 \left(\begin{array}{cc} 0 & \Pi
\\ \Pi^\dagger & 0 \end{array}\right)\, . \ee
This Hamiltonian $H_{AB}$ acts on the spinor $\Psi=\binom{\Psi_1}{\Psi2}$, $\Psi_1$ and
$\Psi_2$ being the wave function at the sites
$A_1$ and $B_2$ respectively. Here $\Pi=p_x+ip_y$ ($p_{x,y}=-i\hbar\partial_{x,y}$) refers to the
in-plane complex momentum operator, $m_* =\frac{\gamma_1}{{2v_0^2}} \approx 0.035
m_e$ is the effective mass in the quadratic dispersion, where $m_e$ is the bare electron mass,
$v_0=\frac{3}{2} \frac{\gamma_0 a}{\hbar}$ is a characteristic velocity (that of massless Dirac fermions in a graphene monolayer), and $v_3=\frac{3}{2} \frac{\gamma_3 a}{\hbar}$ is a velocity accounting for
trigonal warping and which originates from the orbital overlap between $A_1$ and $B_2$ ($a \approx
 0.142$~nm is the interatomic distance). The Hamiltonian in the $K'$ valley is obtained from that in the $K$ valley via the substitution $\Pi\rightarrow -\Pi^\dagger$. In the following, we consider intra-valley effects and for simplicity restrict ourselves to the $K$ valley.

Recent works \cite{{de Gail1},{de Gail2},{Shallcross},{Lopes},{Falko1}} have shown
that a change in the stacking order of the $AB$-bilayer graphene affects drastically the
low-energy electronic structure of the device and may be taken approximatively into
account by adding to the $AB$-bilayer Hamiltonian $H_{AB}$ a
constant energy shift $\Delta$: \be h_\Delta= \left(\begin{array}{cc} 0 &\Delta
\\ \Delta^* & 0 \end{array}\right) \ee which is generically a complex number, which we write:
$$\Delta=\left |\Delta  \right | e^{2i\theta}\, . $$
Below, we briefly review different stacking defaults -- twist, slide, strain, etc -- of the perfect $AB$ bilayer leading to SGB.

References \onlinecite{{Shallcross},{de Gail1},{de Gail2},{Lopes}} have studied the twisted bilayer. They predicted that starting from the perfect $AB$-bilayer graphene and rotating the upper layer by some small
twist angle $\Theta$ with respect to the lower one affects significantly the interlayer hopping $\gamma_1$ (see Fig.\ref{figure1}(b)).
This yields in the reciprocal space a change in the orientations of the Brillouin zones of the two separate monolayers: the Dirac point at $K$ in the lower layer no longer coincides with that at $K_\Theta$ in the rotated upper layer. There is a shift in the momentum space by a wave vector $ \Delta K=2|\vec K|\sin \frac{\Theta}{2}$; where $|\vec K |=\frac{2\pi}{3a}$. In such a case, the energy shift is: $$ \Delta\propto {\left
(\Delta{K}\right)}^2\approx {|\vec K|}^2\Theta^2 \, .$$
In the case of a translational mismatch, Y.W. Son et al. \cite{Son} demonstrated that sliding one graphene layer against the other layer along a vector $\vec{d_s}=(d_x,d_y)$ results in an energy shift:
$$\Delta\propto d_y-id_x\, .$$
For the case of strained graphene, M. Mucha-Kruszynki et al.\cite{Falko1} predicted that a
homogeneous lateral strain applied to the perfect $AB$ graphene bilayer induces two Dirac mini-cones instead of the quadratic low-energy dispersion. Such deformation affects the hopping
parameters $\gamma_0$ and $\gamma_3$ which become dependent on the direction of the strain. Using a tight binding model one finds that the constant energy shift added to $H_{AB}$ has the form
$\Delta=\left |\Delta  \right | e^{2i\theta} $ where $(-\theta)$ is the angle between the
principal axis of the strain tensor and the crystallographic
direction of the crystal and
$\left |\Delta  \right |=(\frac{3}{4})(\eta_3-\eta_0)\gamma_3(\delta-\delta')$ with
$\eta_{0,3}=\frac{d\ln\gamma_{0,3}}{d\ln a}$. Here $\delta$ and
$\delta'$ are the two principal values of the strain tensor.
Eventually, we mention that adding a constant shift to the unperturbed Hamiltonian of BG could also describe the nematic state of an interacting $AB$ graphene bilayer in a mean-field treatment of the spontaneous symmetry breaking \cite{Vafek}.

\begin{figure*}[t]
  \centering
  \begin{tabular}{ccc}
\setlength{\unitlength}{1mm}
\begin{picture}(40,50)
\linethickness{0.5pt}
\put(-40,30){\includegraphics[width=0.25\textwidth]{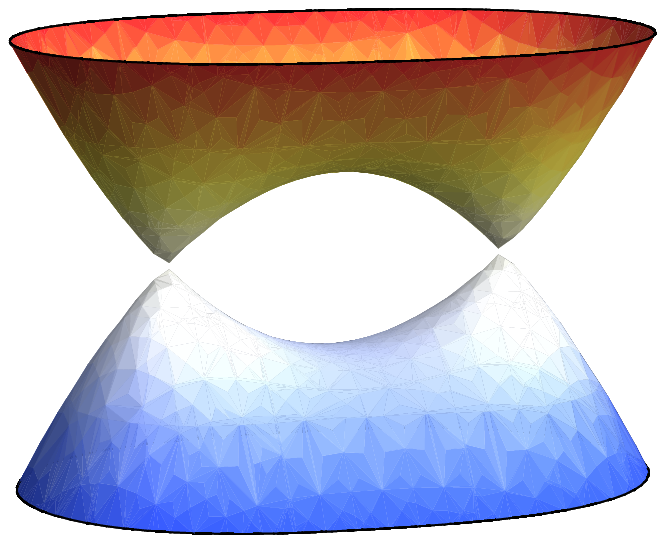}}
\put(0,30){\includegraphics[width=0.25\textwidth]{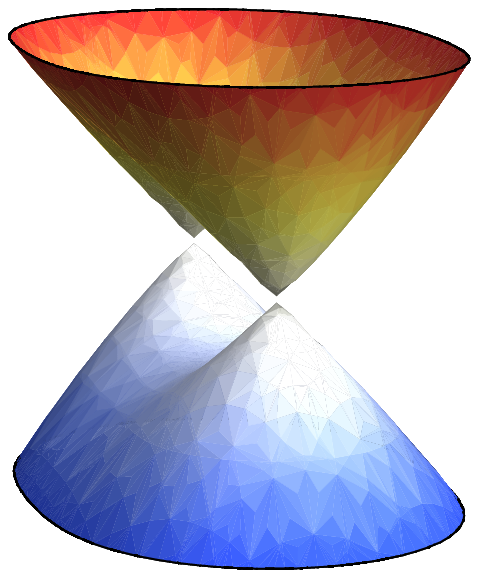}}
\put(40,30){\includegraphics[width=0.25\textwidth]{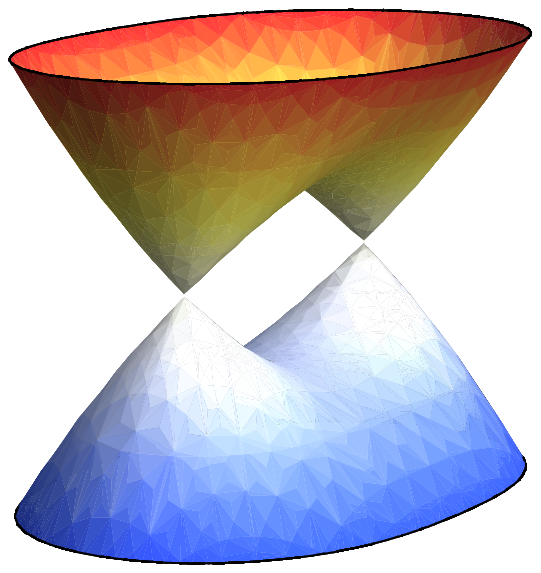}}
\put(-19.5,41.5){\vector(0,1){7}}\put(-19.5,41.5){\vector(0,-1){0.1}}
\put(-23,45){\makebox(0,0){$2\left|\Delta\right|$}}
\put(-50,44){\vector(0,1){10}}\put(-50,44){\vector(1,0){10}}\put(-50,44){\vector(1,1){6}}
\put(-38,44){\makebox(0,0){$k_x$}}\put(-50,55){\makebox(0,0){$\epsilon$}}\put(-42,52){\makebox(0,0){$k_y$}}

\put(-22,-1){\makebox(0,0){(a) $\Delta(\alpha,0)\propto \alpha^2>0$}}\put(15,-1){\makebox(0,0){(b) $\Delta(\alpha,\frac{\pi}{2})\propto -\alpha^2<0$}}\put(55,-1){\makebox(0,0){(c) $\Delta(\alpha,\theta)\propto \alpha^2e^{i2\theta} \in \mathbb{C}$}}
\put(-38,0){\includegraphics[width=0.18\textwidth]{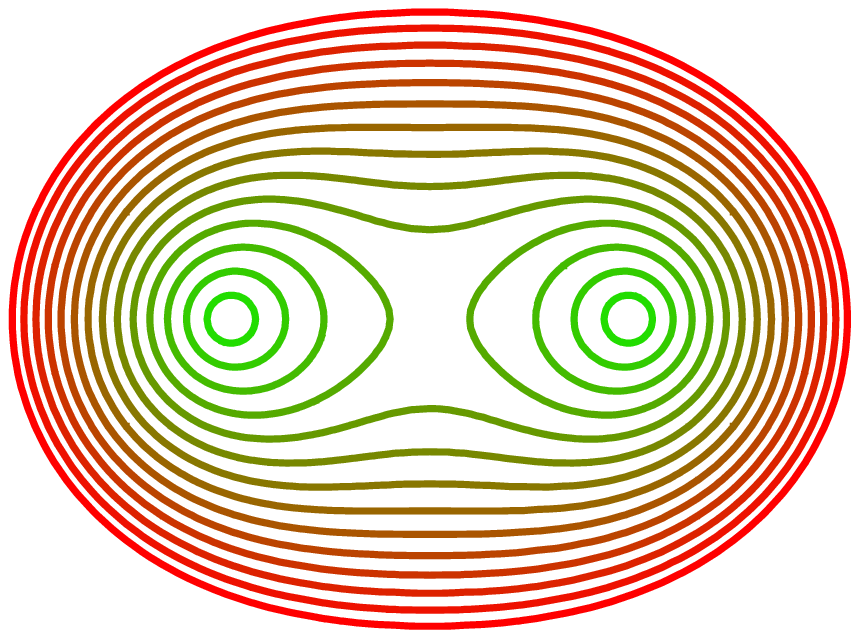}}
\put(-28,16){\circle*{1.5}} \put(-15.5,16){\circle*{1.5}}
\put(2,0){\includegraphics[width=0.18\textwidth]{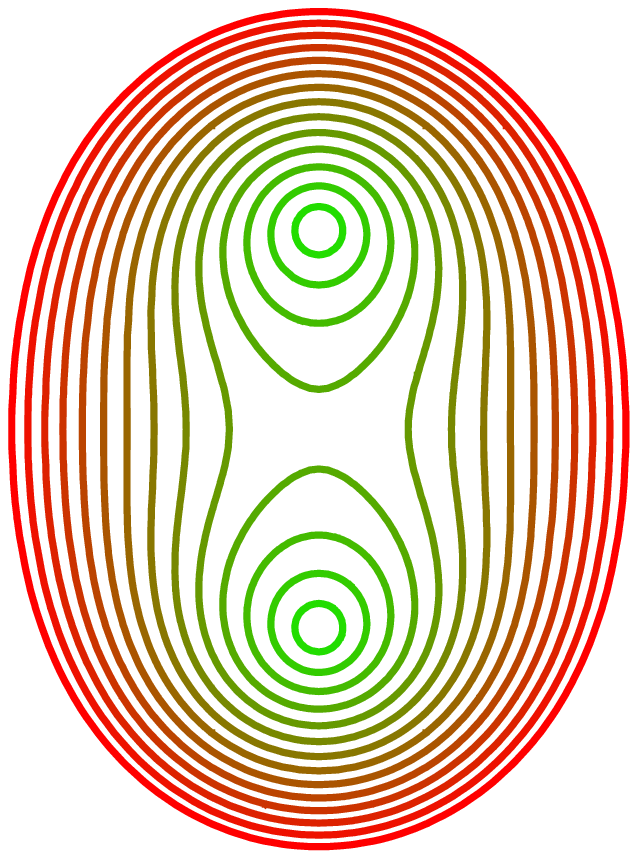}}
\put(18.2,10){\circle*{1.5}} \put(18.2,22){\circle*{1.5}}
\put(42,0){\includegraphics[width=0.18\textwidth]{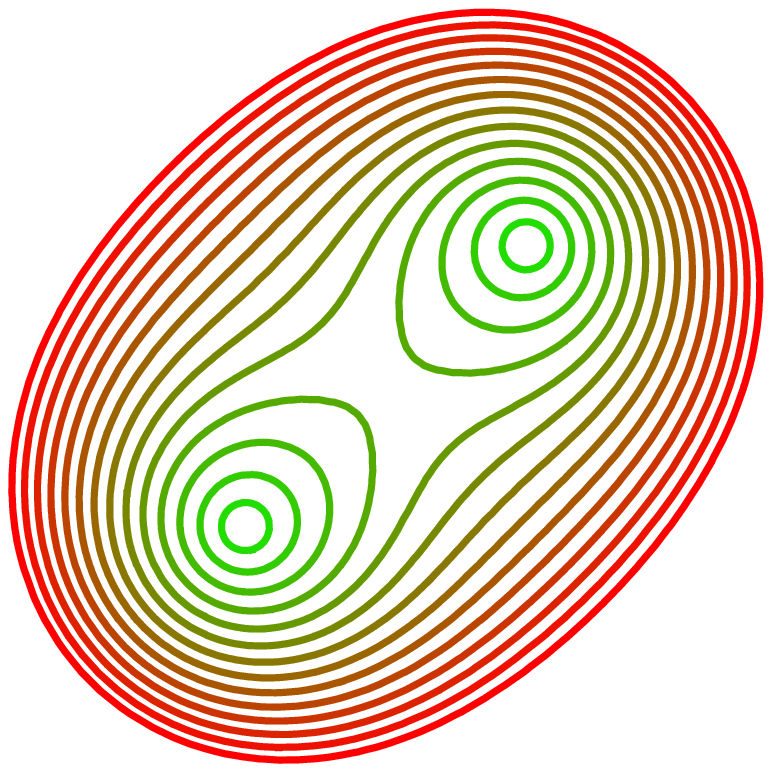}}
\put(54,11.8){\circle*{1.5}} \put(62.5,20.5){\circle*{1.5}}
\put(78,15.5){\vector(1,0){10}}\put(73.5,11){\line(1,1){9.5}}
\put(73,13){\vector(1,1){9}}\put(73,13){\vector(-1,-1){0.1}}
\put(81,16){$)$}\put(80.7,17.5){\vector(-1,-1){0.1}}
\put(73.5,11){\circle*{1.5}} \put(83.2,21){\circle*{1.5}}
\put(90,16){\makebox(0,0){$k_x$}}\put(86,19){\makebox(0,0){$(-\theta)$}}\put(75,19){\makebox(0,0){$2\alpha$}}

\put(-50,16){\vector(0,1){10}}\put(-50,16){\vector(1,0){10}}
\put(-38,16){\makebox(0,0){$k_x$}}\put(-50,28){\makebox(0,0){$k_y$}}
\end{picture}
\end{tabular}
\caption{(Color online) Top: Evolution of the low-energy band structure of the shifted graphene bilayer around the $K$ valley as a function of the orientation of the two Dirac cones as controlled by $\theta$ and for a fixed momentum shift $\alpha$ (the energy shift is $\Delta=\frac{\hbar^2}{2m_*}\alpha^2e^{i2\theta}$): (a) $\theta=0$ (left panel), (b) $\theta=\pi/2$ (central panel) and (c) $0<\theta<\pi/2$ (right panel). Bottom: Corresponding positions of the two Dirac points and iso-energy curves in momentum space $(k_x,k_y)$ close to a single valley.}
\label{figure2}
\end{figure*}

Throughout this work, we restrict ourselves to study deformations (or shifts)
large enough to neglect the trigonal warping term i.e.
\be \frac{\gamma_0 ^2}{\gamma_1}(\frac{ \left | \Delta \right |}{\gamma_3})^2\gg
m_* v_3^2\sim \gamma_1 (\frac{\gamma_3}{\gamma_0})^2 \ee
which means:
\be \left | \Delta \right | \gg \gamma_3
\frac{\gamma_1}{\gamma_0}\, .\ee
Therefore the low energy dispersion of the SGB that we consider in the following
is given by the effective two-band continuum Hamiltonian
 \be \label{effective hamiltonian}
H\simeq h_{AB}+h_{\Delta}\ee
where $\Delta$ is an arbitrary complex energy shift. 
For later convenience, one introduces the parameter $\alpha= \sqrt{\frac{2m_*\left | \Delta \right |}{\hbar^2}}$. In the following, the energy shift $\Delta=\left |\Delta  \right | e^{ 2i\theta}=\frac{\hbar^2}{2m_*}\alpha^2 e^{2 i \theta}$ is parameterized by $(\alpha,\theta)$, where $\alpha$ corresponds to the momentum shift -- it is half the distance between the two Dirac points in momentum space -- and the phase $\theta$ indicates the direction whereby the two Dirac points are separated in momentum space. Varying $\theta$ on $[0,\pi]$, the complex energy shift $\Delta$ sweeps between two limiting values: positive real energy shift (Dirac points separated along $k_x$-direction in momentum space) and negative real energy shift (Dirac points separated along $k_y$-direction in momentum space). 

Solving for the general spectrum of the Hamiltonian in Eq. (\ref{effective hamiltonian}) in the case of an arbitrary complex energy shift $\Delta(\alpha,\theta)$ one finds two
energy bands given by:
\begin{widetext}
\be
\label{dispersion}
\begin{split}
\varepsilon =\pm
\frac{\hbar^2}{2m_*}\sqrt{(k_x^2-k_y^2-\alpha ^2\cos(2\theta))^2+(2k_xk_y+\alpha ^2\sin(2\theta))^2}
\end{split}
\ee
\end{widetext}
The dispersion relation of the SGB (Eq. (\ref{dispersion})) is plotted in Fig. \ref{figure2}. It consists of two Dirac cones separated by $2\alpha$ along the $k_x$-direction in the case of a positive energy shift (see Fig. \ref{figure2}(a)) and along the $k_y$-direction for a negative energy shift (see Fig. \ref{figure2}(b)). In the generic case, the two Dirac cones make an angle $\theta$ with the $k_x$ axis (see Fig. \ref{figure2}(c)).

The location of the two Dirac points and the distance between them in the momentum space are expected to have crucial effect on the transport properties of the shifted bilayer device. In the following section, we address the effect of the shift on the transmission of charge carriers across an undoped graphene bilayer structure connected to reservoirs via interfaces at $x=0$ and $x=L$ (see Fig. \ref{figure3}). The two terminal device is such that the angle $\theta$ controls the orientation of the direction connecting the two Dirac points with respect to the interface between the reservoirs and the ``system'' made of an undoped shifted graphene bilayer. When $\theta=\pi/2$ (resp. $0$) the direction connecting the two Dirac points is parallel (resp. perpendicular) to the interfaces (at $x=0$ and $x=L$).
We discuss separately the case of transmission through potential barrier with Dirac cones in series along the free direction of propagation (which corresponds to $\Delta>0$, i.e. $\theta=0$), lateral Dirac cones (corresponding to $\Delta<0$, i.e. $\theta=\pi/2$) and then the general case of an arbitrary orientation of the two cones ($\Delta \in \mathbb{C}$, i.e. intermediate $\theta$ between $0$ and $\pi/2$).

\section{Transmission probability of a shifted graphene bilayer}
\subsection{Model}
\begin{figure*}[t]
  \centering
\setlength{\unitlength}{1mm}
\begin{picture}(40,50)
\linethickness{0.5pt}
\put(2,10){\includegraphics[width=0.15\textwidth,height=4cm]{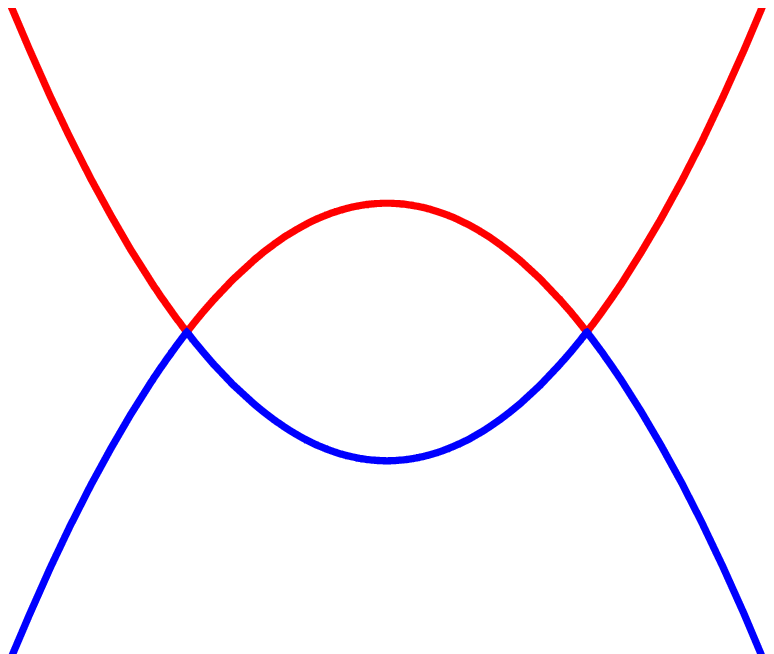}}
\put(0,5){\vector(0,1){40}}\put(-20,30){\vector(1,0){70}}
\put(0,47){\makebox(0,0){$V(x)$}}\put(52,30){\makebox(0,0){$x$}}
\put(2,28){\makebox(0,0){$0$}}\put(32,31){\makebox(-8,-6){$L_x$}} \put(-22,30){\makebox(0,0){$\epsilon_F$}}\put(4,10){\makebox(0,0){$-V_0$}}
\linethickness{1.5pt}
\put(-20,10){\line(1,0){20}}
\put(30,10){\line(1,0){20}}
\put(30,10){\line(0,1){20}}
\put(0,10){\line(0,1){20}}
\put(0,30){\line(1,0){30}}
\end{picture}
\caption {Potential profile in the shifted bilayer device (with $\Delta\propto \alpha^2>0$) in the longitudinal direction $x$. The regions outside $[0,L_x]$ are strongly $n$ doped ($V_0\gg |\Delta|$) and correspond to reservoirs (or leads), whereas the central region is undoped (the Fermi level is at the charge neutrality point, i.e. the Dirac points).}
\label{figure3}
\end{figure*}

In order to explore the effect of an arbitrary shift on electronic transmission in
undoped graphene bilayer, we use a setup similar to that of Refs.
\cite{{Katsnelson_monolayer},{Katsnelson_bilayer},{Snyman},{Tworzydlo}},
sketched in Fig. \ref{figure3}. We consider a two-terminal strip geometry of length $L_x$ along the longitudinal $x$ direction and width $L_y$ along the transversal $y$ direction $(L_x\ll L_y) $, assuming that the sample is
connected at $x=0$ and $x=L_x$ with leads (reservoirs) made out of strongly $n$-doped graphene bilayer. The doping can be thought to be controlled by an external gate which induces a potential profile along the $x$-direction as follows: \be V(x)=\begin{cases}
 &0 \text{ if } 0\leq x\leq L_x\\
 &-V_0 \text{ otherwise}
\end{cases}\ee
where $V_0=-\varepsilon_F\simeq-\frac{\hbar^2 k_F^2}{2m_*}$ when $V_0\gg\left | \Delta \right |$. The potential step is sharp but is assumed to vary over a length scale larger than the in-plane interatomic distance and smaller than the electron wavelength so that it does not induce intervalley scattering of electron across the potential barrier. We assume that an incident electron is coming from negative $x$ and is essentially moving towards positive $x$. To solve the above tunneling problem, we first identify the solutions of the Schr\"odinger equation in the various regions of the device.

Due to the translational invariance along the $y$ direction, we search for solutions proportional to $e^{ik_y y}$ and choose periodic boundary conditions such that $k_y$ is quantized as $k_y=2\pi N/L_y$, $N
= 0, \pm 1, \pm 2, \ldots$. The energy dispersion in the leads is almost parabolic ($V_0\gg\left | \Delta \right |$), so that the leads can be described by heavily doped perfectly stacked $AB$ bilayer, and the shift can be neglected in the reservoirs. In the case of a graphene bilayer, the wave functions outside the potential barrier include both propagating and evanescent waves solutions of the Schr\"odinger equation \cite{{Klein},{Katsnelson_bilayer}}, so that:
\be \left\{\begin{matrix}
\Psi_1(x,y)=(e^{ik_Fx}+re^{-ik_Fx}+Ee^{k_Fx})e^{ik_yy}\\\Psi_2(x,y)=(e^{ik_Fx}+re^{-ik_Fx}-Ee^{k_Fx})e^{ik_yy}
\end{matrix}\right.\ee for $x< 0$
\be
\left\{\begin{matrix}\Psi_1(x,y)=(te^{ik_F(x-L_x)}+Fe^{-k_F(x-L_x)})e^{ik_yy}
\\ \Psi_2(x,y)=(te^{ik_F(x-L_x)}-Fe^{-k_F(x-L_x)})e^{ik_yy}
\end{matrix}\right.
\ee for $x>L_x$, where $r$, $t$, $E$ and $F$ are yet unknown complex coefficients.
To identify the wave functions inside the potential barrier $(0 < x
<L_x)$ where the shifted bilayer is undoped, we solve
the eigenvalue equation with zero energy $H\Psi=0$ .
The two components of $\Psi$ satisfy the following equations:
\be \label{system_psi}
 \left\{\begin{matrix}[(\frac{\partial }{\partial
x}+i\frac{\partial }{\partial
y})^2-\alpha e^{-i\theta}]\Psi_1=0
\\ [(\frac{\partial }{\partial
x}-i\frac{\partial }{\partial
y})^2-\alpha e^{i\theta}]\Psi_2=0\end{matrix}\right. \ee
For an arbitrary complex energy shift $\Delta(\alpha,\theta)$, the general solutions of these equations is given by:
\begin{widetext}\be \label{wavefunction} \begin{split}
 \left\{\begin{matrix} \Psi
_1(x,y)=[A\exp(k_yx+ie^{-i\theta}\alpha
x)+B\exp(k_yx-ie^{-i\theta}\alpha x)]e^{ik_yy}
\\\Psi
_2(x,y)=[C\exp(-k_yx+ie^{i\theta}\alpha
x)+D\exp(-k_yx-ie^{i\theta}\alpha x)]e^{ik_yy}
\end{matrix}\right.
\end{split}
\ee
\end{widetext}
Here $A$,$B$,$C$,$D$,$E$ and $F$ are complex constants, $r$ is the reflection amplitude and $t$
the transmission amplitude. The two limits $\theta=0$ and $\theta=\frac{\pi}{2}$ correspond to positive $(\Delta > 0)$ and negative $(\Delta < 0)$ shift, respectively. Note that, when $\theta=0$, the wave function in Eq. (\ref{wavefunction}) is a mixture of $\alpha$-oscillating and $k_y$-evanescent waves, whereas, for $\theta=\frac{\pi}{2}$ the solutions in Eq. (\ref{wavefunction}) contain only evanescent waves.

To find the transmission coefficient through barrier, we require the continuity of both components of the wave function and their derivatives at $x = 0$ and $x =L_x$ , which yields eight equations for the eight unknown coefficients: \be \label{system_t} \left\{\begin{matrix}
1+r+E=A+B\\1+r-E=C+D
\\ i-ir+E=a_1A+b_1B
\\ i-ir-E=c_1C+d_1D
\\t+F=a_2A+b_2B
\\t-F=c_2C+d_2D
\\it-F=a_1a_2A+b_1b_2B
\\ it+F=c_1c_2C+d_1d_2D
\end{matrix}\right.
\ee
where $a_1=\frac{k_y+ie^{-i\theta}\alpha}{k_F}$,
$b_1=\frac{k_y-ie^{-i\theta}\alpha}{k_F}$,
$c_1=\frac{-k_y+ie^{i\theta}\alpha}{k_F}$,
$d_1=\frac{-k_y-ie^{i\theta}\alpha}{k_F}$ and $a_2=\exp (k_FL_xa_1)$, $b_2=\exp (k_FL_xb_1)$, $c_2=\exp (k_FL_xc_1)$, $d_2=\exp (k_FL_xd_1)$.
We solve these equations for the transmission amplitude $t$ (see below). The total transmission probability $T$ can be directly found from the transmission amplitude using the relation $T =\left | t\right |^2$.

In the following sections, we focus on the effect of the incidence angle (as controlled by $k_y$) and the stacking defect (as parameterized by $\alpha$ and $\theta$) on the zero-energy transmission probability. We first study the situation where the two Dirac cones are hit in succession ($\theta=0$), then when they are approached in parallel ($\theta=\pi/2$) and finally consider an intermediate case ($0<\theta<\pi/2$).

\subsection{Transmission for Dirac cones in succession}
\begin{figure}[h]
\includegraphics[width=0.45\textwidth]{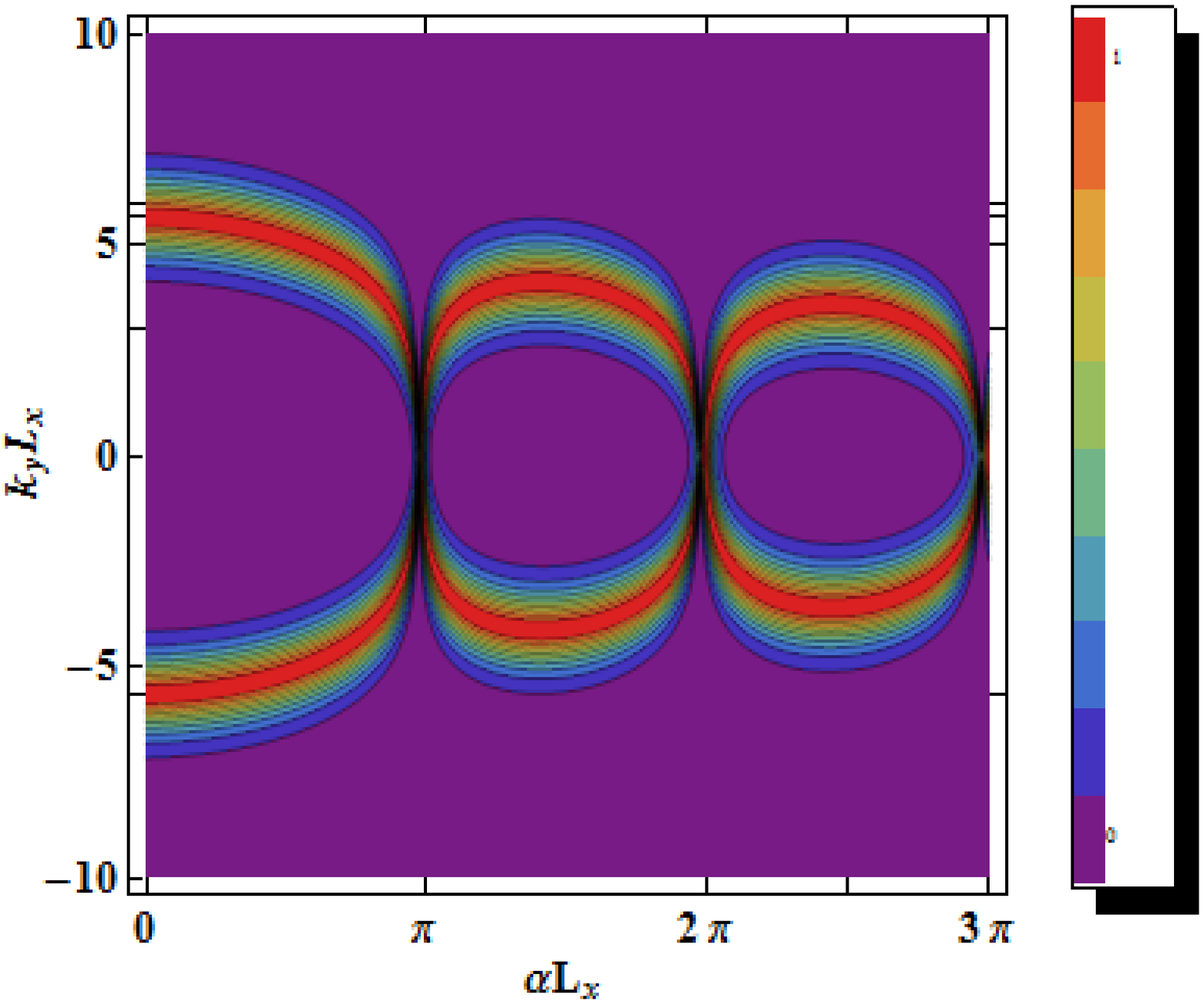}
\caption {(Color online) Contour plot of the transmission probability $T$ of the shifted
bilayer with $\Delta(\alpha,\theta=0)$ as a function of the momentum shift $\alpha L_x$ and of the perpendicular momentum $k_yL_x$. The doping of the reservoirs is taken such that $k_FL_x=200$.}
\label{figure4}
\end{figure}

In this Section we present our main result, namely the transmission across an undoped SGB strip in the case of a stacking defect defined by a positive energy shift $\Delta (\alpha,0)$, which corresponds to the two Dirac points being hit in succession by an incident electron (see Fig. \ref{figure2}(a)). The linear system of equations (\ref{system_t}) is solved for the transmission amplitude $t$ which is found to be:
\begin{widetext}
\be
\begin{split}
t=\frac{2ik_FL_x\textrm{sinc}(\alpha
L_x)\cosh{(k_yL_x)}[1-(\frac{\alpha}{k_F})^2-(\frac{k_y}{k_F})^2+2\frac{\cos{(\alpha
L_x)}}{k_FL_x \textrm{sinc}(\alpha L_x)}]}{2i \cos^2{(\alpha
L_x)}-\frac{(k_y^2+\alpha ^2+ik_F^2)^2 L_x^2}{k_F^2} \textrm{sinc}^2(\alpha
L_x)+2(1-i)\frac{(k_y^2+\alpha ^2+ik_F^2)L_x}{k_F} \textrm{sinc}(2\alpha L_x)+2i
\cosh^2{(k_y L_x)}}
\end{split}
\label{transmissionamplitude}
\ee
\end{widetext}
The transmission amplitude $t$ remains invariant under $k_y\rightarrow-k_y$ meaning that the
transmission probability $T=|t|^2$ is symmetric with respect to normal incidence. Figure \ref{figure4} shows a contour plot of the transmission probability $T=|t|^2$ as a function of the momentum shift
$\alpha L_x$ and the transverse momentum $k_yL_x$ (controlling the incidence angle in the vicinity of normal incidence).

\begin{figure}[h]

\includegraphics[width=0.45\textwidth]{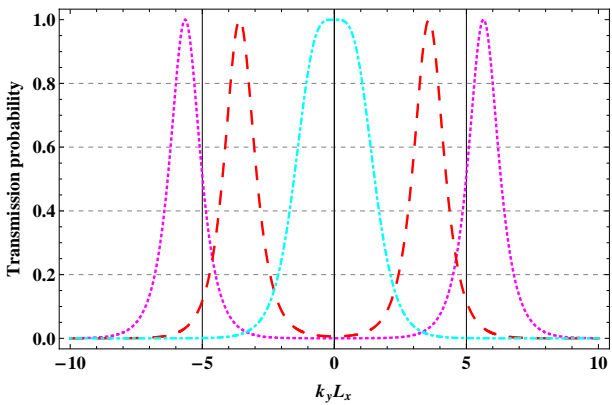}
\caption{ (Color online) Transmission probability of the shifted
bilayer with $\Delta(\alpha,\theta=0)$ as a function of the perpendicular momentum
$k_yL_x$ for three different values of the momentum shift $\alpha L_x=0$ (dotted
magenta), $\pi$ (dot-dashed cyan) and $5\pi/2$ (dashed red). The
doping of the reservoirs is taken such that $k_F L_x=200$.}
\label{figure5}
\end{figure}

\begin{figure}[h]

\includegraphics[width=0.45\textwidth]{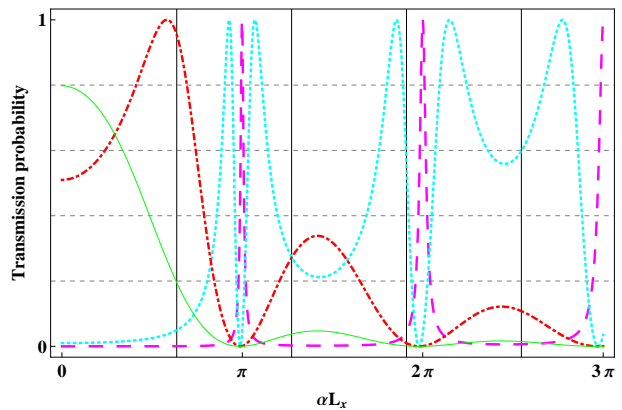}
\caption{(Color online) Transmission probability of the shifted
bilayer with $\Delta(\alpha,\theta=0)$ as a function of the shift $\alpha L_x$
for four fixed values of the perpendicular momentum: $k_yL_x=0$ (large-dashing
magenta), $3$ (dotted cyan), $5$ (dot-dashed red) and $6$ (thick green). The
doping of the reservoirs is taken such that $k_F L_x=200$.}
\label{figure6}
\end{figure}

For a given transverse momentum $k_y$, the transmission $T$ is an oscillating function of the momentum shift $\alpha$ with a period $\pi/L_x$. Examples of cuts in the $(\alpha L_x,k_yL_x)$-contour plot of $T$ along different fixed values of $\alpha$ momentum then along different fixed values of $k_y$ momentum are plotted in Fig. \ref{figure5} and Fig. \ref{figure6}.

From Fig. \ref{figure5}, one can see that for a given momentum shift $\alpha$ the transmission probability
exhibits pronounced peaks at some transverse momentum $k_y$ or equivalently at
some incidence angles, where the barrier becomes perfectly transparent. These peaks are transmission resonances. The transmission $T$ depends sensitively not only on the incidence
angle but also on the momentum shift: the form, the position and the
number of resonance peaks vary with the stacking configuration of
the bilayer graphene device.
When looking at the form and the number of peaks (see Fig. \ref{figure5}), one can
distinguish two kinds of resonant behavior for the transmission
probability $T$ depending on whether the dimensionless momentum shift is quantized as $\alpha L_x= m\pi$,  where $m$ is a positive integer, or not $\alpha L_x\neq m\pi$.
Starting from a stacking configuration where the momentum shift
$\alpha L_x=m\pi $, as shown in Fig. \ref{figure4}, there is one wide resonance
peak which occurs exactly at normal incidence i.e $k_y=0$.
However, this unique resonance peak splits into two narrow peaks as
the momentum shift increases and differs from $m\pi $. These resonance peaks first separate when increasing $\alpha L_x$ before approaching each other and meeting again at normal incidence when $\alpha L_x=(m+1)\pi $.

In Fig. \ref{figure6}, we plot the variation of the transmission probability in terms of the momentum shift  $\alpha L_x$ for fixed values of $k_yL_x$. It is obvious that one should distinguish two kinds of transmission peaks, that are of different nature. The first kind occurs at normal incidence $k_yL_x=0$ (dashed large magenta in Fig. \ref{figure5}), has a $\pi$-periodicity in $\alpha L_x$ and features narrow resonances of perfect transmission. The second kind occurs at oblique incidence (dotted cyan and dot-dashed red in Fig. \ref{figure6}), do not have a $\pi$ periodicity, correspond to wider resonances of maximum probability. Also, depending on the incidence angle, the number of resonance peaks with unit transmission probability varies. Beyond a critical value of $k_yL_x$, i.e incidence angle, the transmission probability decreases and transmission resonances do not reach $T=1$ anymore (thick green in Fig. \ref{figure6}). Eventually, for even larger incidence angles, particles are totally reflected (see Fig. \ref{figure4} and Fig. \ref{figure6}). Thus, for fixed incidence angles, transmission resonances could be controlled by tailoring the stacking defect.

\begin{figure}[h]

\includegraphics[width=0.45\textwidth]{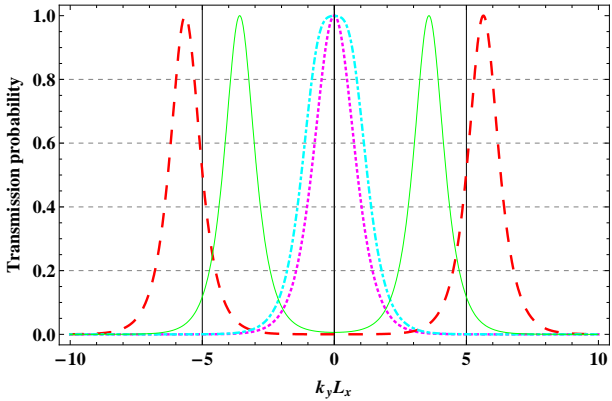}
\caption{(Color online) Comparison of the transmission probability as a function of the perpendicular momentum $k_yL_x$ for the shifted bilayer with $\Delta(\alpha,\theta=0)$ (thick green for $\alpha L_x=\frac{5\pi}{2}$ and dot-dashed cyan for $\alpha L_x=\pi$), the monolayer \cite{Katsnelson_monolayer}(dotted magenta) and the perfect $AB$ bilayer\cite{Katsnelson_bilayer} (large-dashing red, corresponding to $\alpha L_x=0$). The
doping of the reservoirs is taken such that $k_F L_x=200$.}
\label{figure7}
\end{figure}

To elucidate the effect of the shift on the transmission resonances across undoped
graphene bilayer and to give a deeper understanding of the origin of these resonances,
it is instructive to compare the angular dependence of SGB to that of $AB$-
bilayer\cite{Katsnelson_bilayer}.
Accordingly, we consider the limit of highly doped leads ($k_F\gg
k_y$,$\alpha$,$L_x^{-1}$) and a
straightforward algebra yields to the analytical expression of the
transmission amplitude $t$:
 \be \label{t approx}
t=\left\{\begin{matrix} \frac{2ik_FL_x\cosh(k_yL_x)
\textrm{sinc}(\alpha L_x)}{k_F^2L_x^2 \textrm{sinc}^2(\alpha L_x)+2i\cosh^2
(k_yL_x)} \textrm{ for }\alpha L_x\neq m\pi
\\ (-1)^m \frac{2\cosh(k_yL_x) }{\cosh^2(k_yL_x)+1}\textrm{ for } \alpha
L_x= m\pi
\end{matrix}\right.
\ee
Note that the transmission amplitude corresponding to a vanishing shift, i.e $\alpha
L_x\rightarrow0$ in the first line, reproduces the one obtained in Ref. \onlinecite{Katsnelson_bilayer}. The resonance occurs at a finite incidence given by  $k_yL_x \simeq \ln( \sqrt{2}k_FL_x)$ (in a monolayer, resonance occurs at normal incidence $k_y=0$). A comparaison of the $k_yL_x$ dependence of the transmission probability across the SGB strip with the perfect $AB$ graphene bilayer and the monolayer is depicted in Fig. \ref{figure7} (the three systems are assumed to be undoped). The location of the transmission resonance across the SGB is controlled by the momentum shift $\alpha$ (see below). In particular for a SGB with $\alpha L_x=m \pi$, there is a unique transmission resonance at normal incidence, which contrasts strongly with the $AB$ bilayer. In addition the angular dependence $T(k_y)$ close to $k_y=0$ is also different from the monolayer (see Fig. \ref{figure7}).

Equations (\ref{t approx}) shows that the transmission $T=1$
occurs at: \be \label{resonance positions} k_yL_x\simeq\left\{\begin{matrix}
 \ln (\sqrt{2}k_FL_x\textrm{sinc} (\alpha L_x)) \textrm{ for } \alpha L_x\neq m\pi\\
0  \textrm{ for } \alpha L_x=m\pi\end{matrix}\right.\ee
This allows one to get some insight into the two distinct kinds of resonances, which can be understood as follows. For a stacking configuration of the SGB where $\alpha L_x\neq m\pi$, the zero-energy transmission resonances occur at  oblique incidence:  $k_yL_x\simeq \ln( \sqrt{2}k_FL_x\textrm{sinc}(\alpha
L_x))$. Adjusting the value of the momentum shift $\alpha$, the incidence angle of the resonances moves closer to normal incidence. These tunable resonances are due to evanescent waves (quantum tunneling) and are related to the Klein tunnel effect. They are also similar to those observed in the undoped perfect $AB$-bilayer graphene where the resonant transmission direction is given by\cite{Katsnelson_bilayer} $k_yL_x
\simeq \ln( \sqrt{2}k_FL_x)$, which indeed corresponds to Eq. (\ref{resonance positions}) when $\alpha=0$.

If $\alpha L_x=m\pi$, with $m$ a positive integer, perfect transmission occurs at normal incidence, i.e $k_y=0$. This is a surprising feature of normally incident chiral fermions across undoped (zero energy) bilayer graphene structure. Indeed in $AB$ stacked bilayer, normal incidence corresponds to perfect {\it reflection} \cite{Klein}. These resonances behave differently from all other known transmission resonances in graphene devices. They occur at zero energy $\varepsilon=0$, for a quantized momentum shift $\alpha L_x=m\pi$ and at normal incidence $k_y=0$. Injecting these three relations in the dispersion relation Eq. (\ref{dispersion}) for $\theta=0$ gives the following resonance condition: $$k_x L_x=m \pi \, .$$ This is just the familiar condition on the wavelength $\lambda=2 L_x/m$ usually associated to Fabry-P\'erot resonances. These resonances originate from the constructive interference of the oscillating waves inside the barrier in between the two reservoirs. The potential barrier can be assimilated to a cavity made of two interfaces (at $x=0$ and at $x=L_x$) as in the well-known optical Fabry-P\'erot interferometer.

In previous theoretical works on resonant transmission at zero-energy in graphene structures, evanescent waves resonances have been found to occur at normal incidence for the monolayer and at oblique incidence for the bilayer. Whereas Fabry-P\'erot resonances were found for non zero energy and at oblique incidence in both structures. The observation of Fabry-P\'erot resonance {\it at zero energy and normal incidence} constitute a striking effect of the shift on the bilayer graphene and the main result of the present paper. Note, in particular, that in a graphene monolayer, Fabry-P\'erot resonances can not occur at normal incidence due to Klein tunneling, which prevents the multiple reflections needed in such an interferometer.

\subsection{Transmission for Dirac cones in parallel}
We now consider zero-energy transmission across SGB with lateral Dirac cones corresponding to $\theta=\pi/2$ (see Fig. \ref{figure2}(b)). Solving the set of equations (\ref{system_t}) for $\Delta(\alpha,\frac{\pi}{2})$, one gets the following expression for the transmission amplitude:
\begin{widetext}
\be
\begin{split}
t=\frac{2ik_FL_x\textrm{sinhc}(\alpha
L_x)\cosh{(k_yL_x)}[1-(\frac{\alpha}{k_F})^2-(\frac{k_y}{k_F})^2+2\frac{\cos{(\alpha
L_x)}}{k_FL_x \textrm{sinhc}(\alpha L_x)}]}{2i\cos^2{(\alpha
L_x)}-\frac{(k_y^2+\alpha ^2+ik_F^2)^2 L_x^2}{k_F^2} \textrm{sinhc}^2(\alpha
L_x)+2(1-i)\frac{(k_y^2+\alpha ^2+ik_F^2)L_x}{k_F}\textrm{sinhc}(2\alpha L_x)+2i
\cosh^2{(k_y L_x)}}
\end{split}
\ee
\end{widetext}
where we defined $\textrm{sinhc}(\alpha L_x)\equiv \frac{\sinh{(\alpha L_x})}{\alpha L_x}$. Note that this equation is similar to Eq. (\ref{transmissionamplitude}) if one substitutes $\textrm{sinc}$ by $\textrm{sinhc}$.

\begin{figure}[h]

\includegraphics[width=0.45\textwidth]{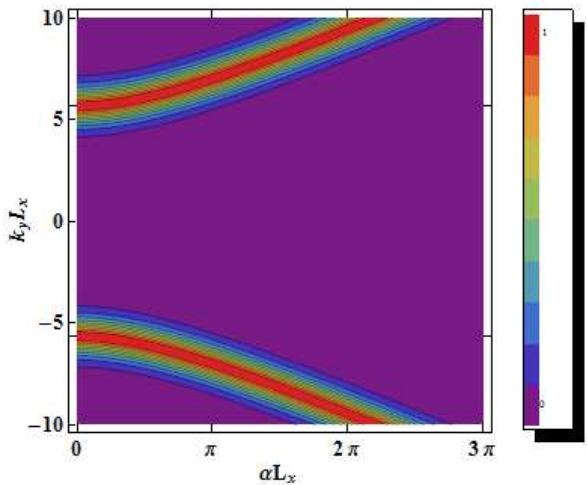}
\caption{(Color online) Contour plot of the transmission probability of the shifted bilayer with $\Delta(\alpha,\theta=\frac{\pi}{2})$ as a function of the momentum shift $\alpha L_x$ and the transverse momentum $k_y L_x$. The doping of the reservoirs is taken such that $k_F L_x=200$.}
\label{figure8}
\end{figure}
Figure \ref{figure8} shows a contour plot of the transmission probability $T=\left |t   \right |^2$ as a function of $\alpha L_x$ and $k_yL_x$. There are again two symmetric $k_yL_x$ directions for which the barrier is transparent. These are the two peaks of the evanescent transmission resonances already encountered in the case of Dirac cones in succession (see previous section), with the difference that their angular dependence is monotonic and does not feature oscillations (compare Fig. \ref{figure4} and Fig. \ref{figure6}). Also, no perfect transmission is found for normally-incident particles, i.e electrons are always perfectly reflected for normal incidence (see Fig. \ref{figure8}). In the limit of highly doped leads, the transmission coefficient across shifted graphene bilayer with lateral Dirac cones takes the following  expression:
\be
t=\frac{2ik_FL_x\cosh{(k_yL_x)} \textrm{sinhc}(\alpha
L_x)}{k_F^2L_x^2\textrm{sinhc}^2(\alpha L_x)+2i\cosh^2 {(k_yL_x)}} \, .\ee
Perfect transmission occurs for:
\be k_yL_x\simeq \ln (\sqrt{2}k_FL_x\textrm{sinhc}(\alpha L_x))\ee
This equation shows that the $k_y$-resonance directions are tunable by the shift $\alpha$. However the angular positions of the resonance move away one from the other as the shift increases. The resonances found correspond to Klein-like tunneling of massive chiral fermions and are due to evanescent waves, as in the case of $AB$ bilayer.

\subsection{Transmission for an arbitrary orientation of the Dirac cones}

\begin{figure*}[t]
  \centering

\setlength{\unitlength}{1mm}
\begin{picture}(40,50)
\linethickness{0.5pt}
\put(-80,0){\includegraphics[width=0.25\textwidth]{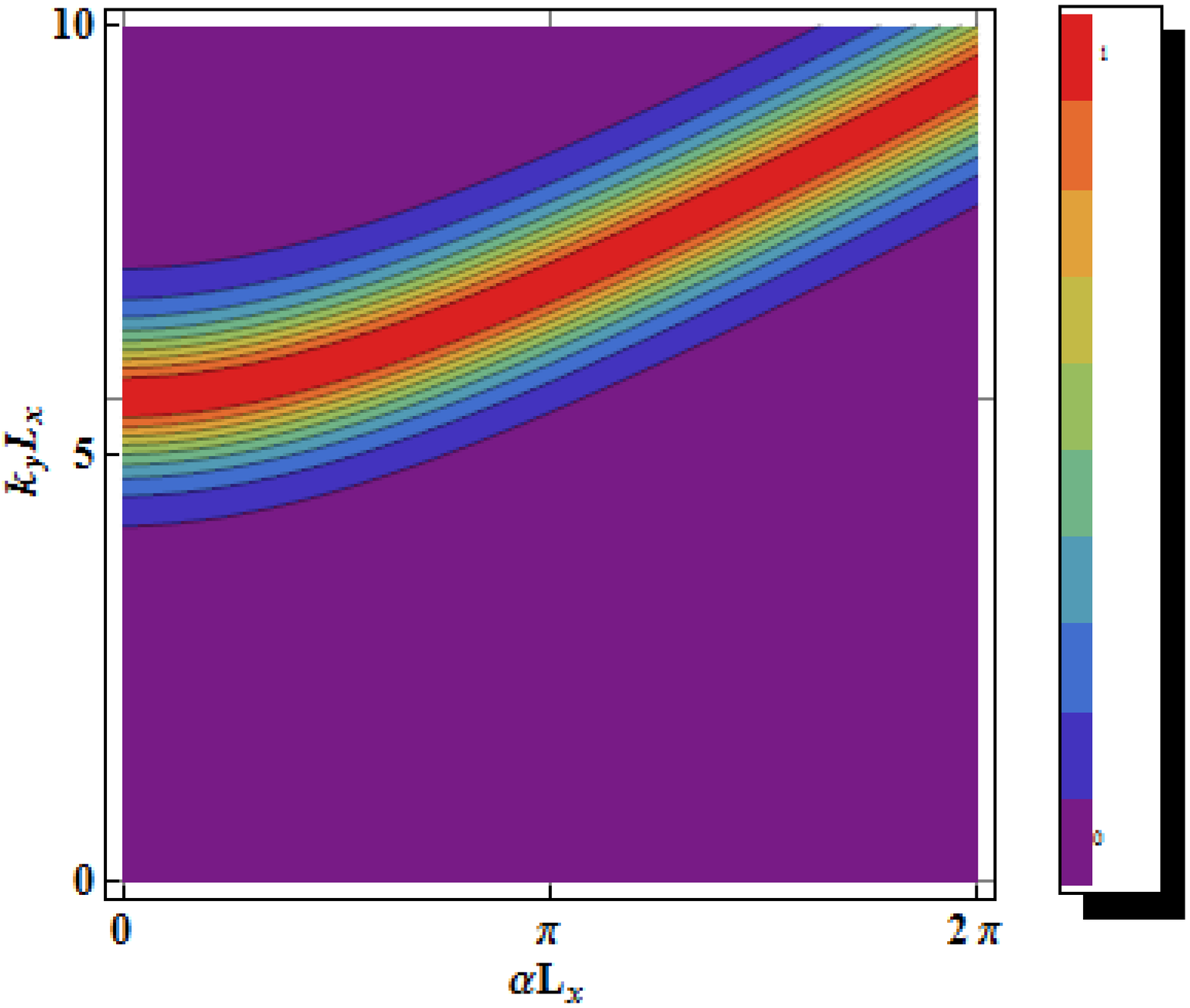}}
\put(-30,0){\includegraphics[width=0.25\textwidth]{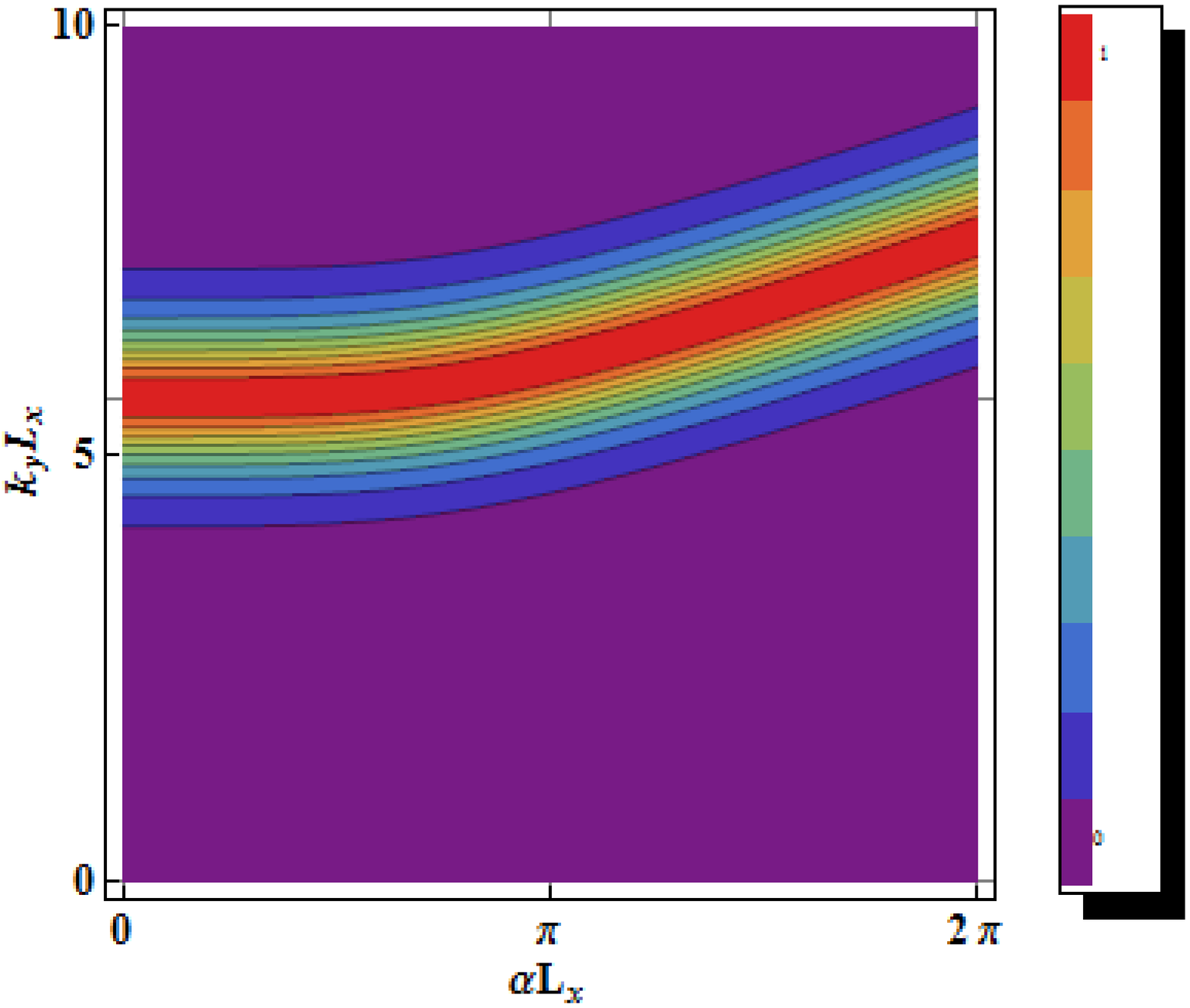}}
\put(20,0){\includegraphics[width=0.25\textwidth]{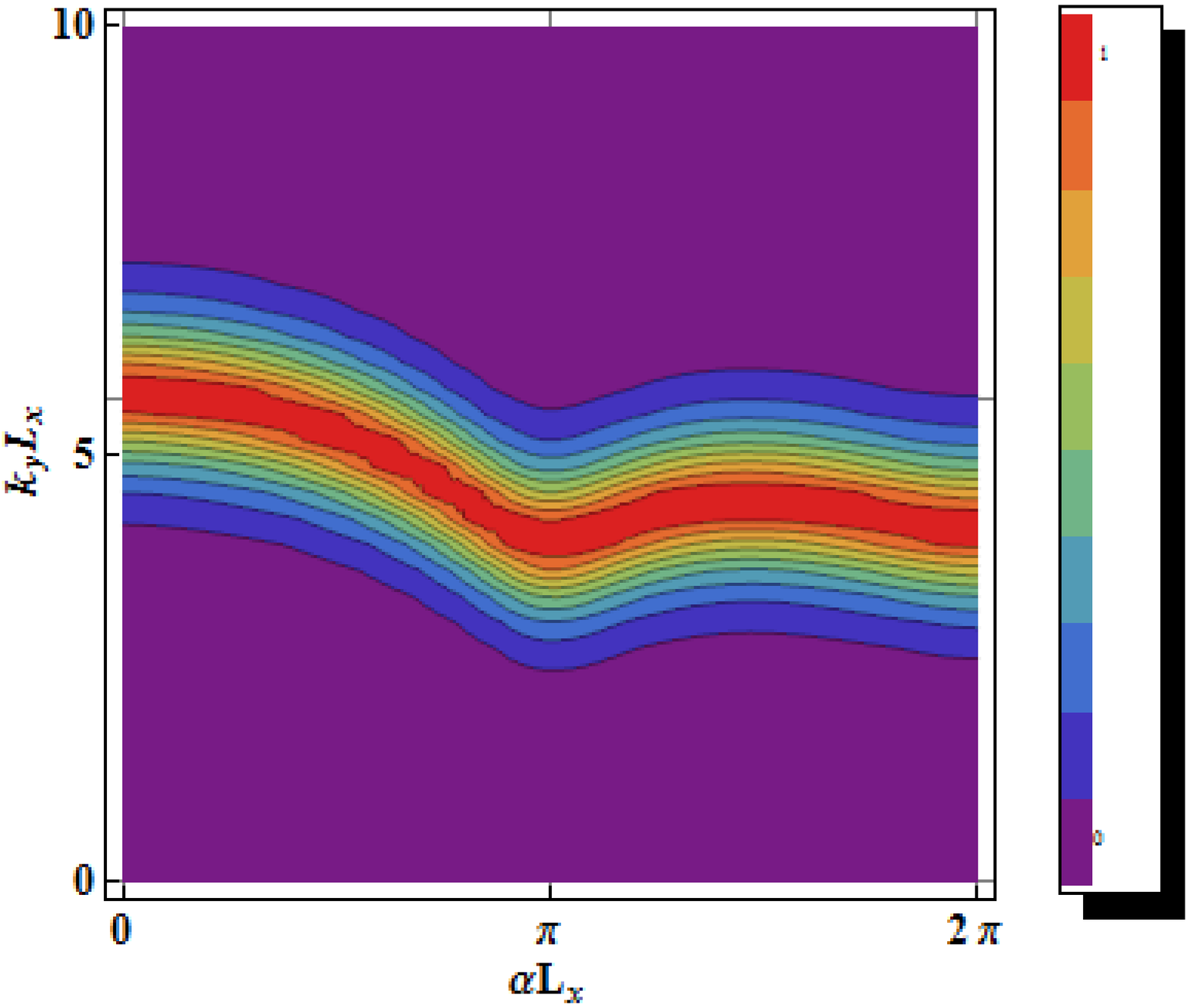}}
\put(70,0){\includegraphics[width=0.25\textwidth]{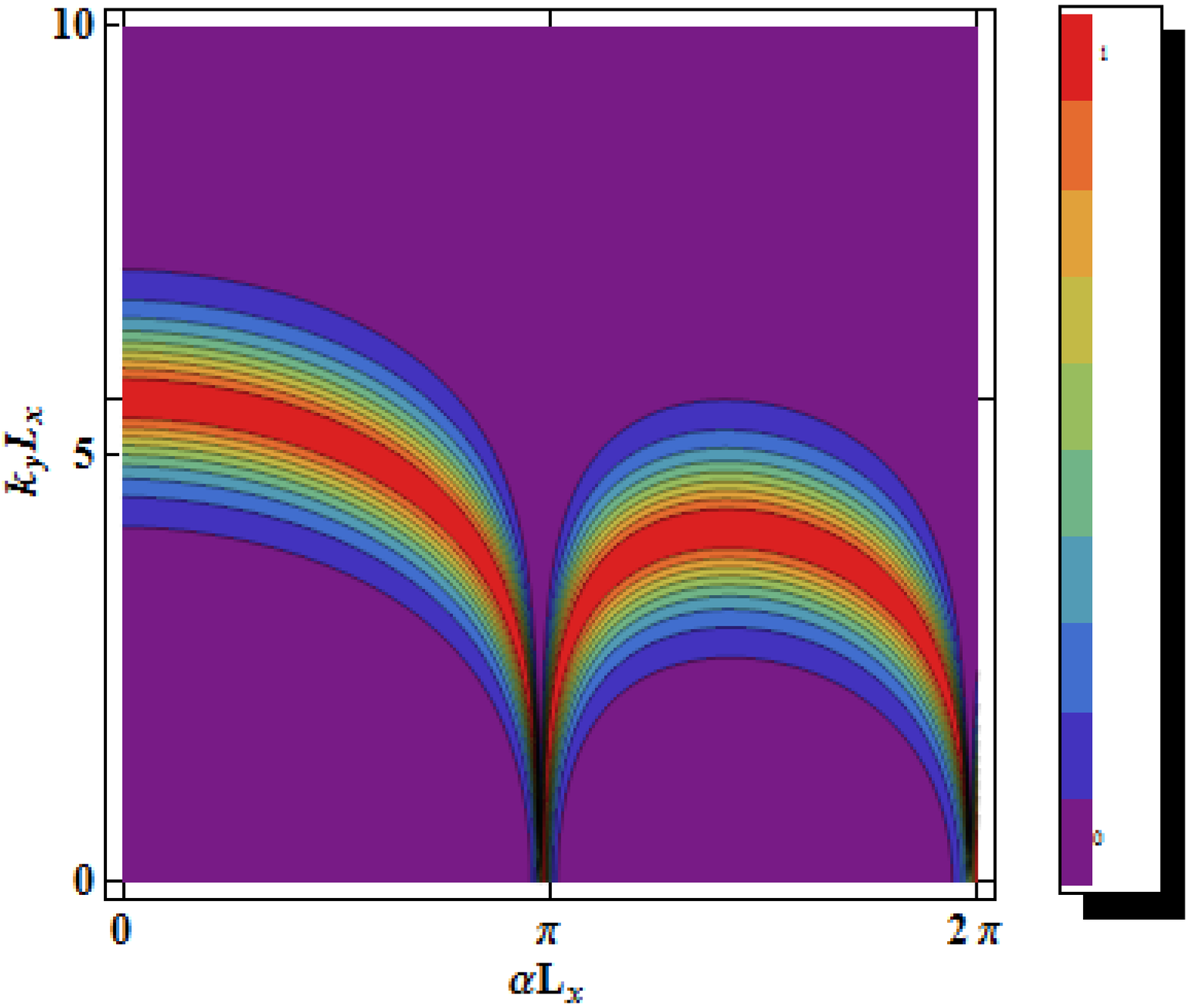}}
\put(-60,45){\makebox(0,0){(a) $\Delta(\alpha,\theta=\frac{\pi}{2})=-\alpha^2$}}
\put(-10,45){\makebox(0,0){(b) $\Delta(\alpha,\theta=\frac{\pi}{4})=i\alpha^2$}}
\put(40,45){\makebox(0,0){(c) $\Delta(\alpha,\theta=\frac{\pi}{18})=e^{i \pi/9}\alpha^2$}}
\put(90,45){\makebox(0,0){(d) $\Delta(\alpha,\theta=0)=\alpha^2$}}

\end{picture}
\caption{(Color online) Transmission probability $T$ (shown as a color plot) as a function of the momentum shift $\alpha L_x$ and the transverse momentum $k_yL_x$ (controlling the incidence angle) for four orientations of the Dirac cones: (a) $\theta=\pi/2$ (lateral Dirac cones), (b) $\theta=\pi/4$, (c) $\theta=\pi/18$ and (d) $\theta=0$ (Dirac cones in succession).}
\label{figure9}
\end{figure*}

For completeness, we give numerical results for zero-energy transmission probability across SGB in the case of an arbitrary complex energy shift. We show in Fig. \ref{figure9} examples of the $(\alpha L_x,k_yL_x)$ contour plot of the transmission probability for four different shift energies: $\Delta (\alpha,\frac{\pi}{2})$, $\Delta (\alpha,\frac{\pi}{4})$, $\Delta (\alpha,\frac{\pi}{18})$ and $\Delta (\alpha,0)$. This figure summarizes all the possible effects of the shift on the transmission through the bilayer structure. As we can see, besides the momentum shift $\alpha$ (the $k$-space distance between the two Dirac points), the position of the Dirac points in the momentum space, given by the parameter $\theta$, plays an important role in the control of the transmission probability. The case $\theta=0$ stands apart: two kinds of resonances are possible (evanescent waves or Fabry-P\'erot) depending on the value of the momentum shift $\alpha$. As soon as $\theta \neq 0$, only one kind of transmission resonances (evanescent waves) is observed.

\section{Experimental implication}
We turn to the discussion of the possible experimental realization of our predictions. The zero-energy transmission resonances in the SGB structure is found to be tuned by an interplay between the stacking defect parameters ($\alpha, \theta$) and the incidence angle ($k_y$). If the stacking of the two layers can be controlled, it means that quantum transport properties become tunable via a new kind of external parameter.

As for monitoring the stacking configuration of the bilayer structure, we mention that producing large scale graphene with controllable number of layers and stacking order has been extensively studied. For instance, in the case of rotationnally faulted bilayer graphene, diverse twisting angles have been observed in existing experiments:  in Ref. \onlinecite{Luican} the twisting angle is around $21.8^{\circ}$ or $3.5^{\circ}$. Among the already available experimental results, L. Brown et al. have reported in Ref. \onlinecite{Brown} twisting angles of CVD grown graphene tri- and bilayer around $29^{\circ}$, $24^{\circ}$, $17^{\circ}$, $12^{\circ}$ and $5^{\circ}$. Such twist angle distribution has been understood in terms of an angle-dependent interlayer potential. The latter, together with superlubricity in twisted bilayer, suggested a theoretical model which will be helpful in providing guidelines for controlling the coverage distribution of bilayer graphene with known twist angle by modifying the thermodynamic driving force and kinetics during the growth \cite{Brown}. 

Now, we move to the discussion of the possible observation of the angle-dependent transmission. Shortly after the theoretical prediction of Klein tunneling \cite{Klein}, several experiments aimed at measuring some of the predicted exotic features of electrons transmission across graphene based devices (such as unusual lensing, Klein tunneling, etc.). For example, Fabry-P\'erot interferences at finite energy have been observed in conductance measurements in both monolayer\cite{Rickhaus,Young,Cho} and bilayer\cite{Cho} devices. Signatures of the preferential transmission of normally incident carriers have been observed indirectly in resistance measurements typically done using devices with two-terminal geometries (see for example Ref. \onlinecite{Huard}). Measuring magneto-conductance across a potential barrier, Young and Kim\cite{Young} were able to detect a sudden shift in Fabry-P\'erot interferences occurring at a critical magnetic field, which can be attributed to Klein tunneling. However, the difficulty of explicitly measuring the angle-dependent transmission remains. Indeed, in the two-terminal geometry of most of these transport experiments -- imposed by the small sample size required to reach the ballistic regime --, the conductance is obtained as an integral over the angular dependence of the transmission probability and therefore does not give a direct information on the angle-dependent transmission.

Very recently, a direct but partial measurement of the angular dependence of the transmission of Dirac electrons across p-n junctions has been reported\cite{Rahman}. Ballistic p-n junction devices with three arms were used: straight and $45^{\circ}$ angled. The barrier height was controlled by a common top-gated portion shared by the three arms so any gate-dependent angle insensitive series resistances affects equally all arms. Using a balancing measurement technique, the angle-dependent effect was separated out from other angle insensitive gate dependent and device-dependent effects. Large fluctuations in the resistance measurements arising from Klein tunneling as a function of gate voltage at $45^{\circ}$ angle as compared to normal incidence have been observed. It has been shown that the excess of fluctuation in the resistance measurements is not other than the contributions from the angle-dependent part of the resistance.  Another partial signature of Klein tunneling was obtained in Ref. \onlinecite{Rickhaus} in the form of the identification of a strong collimation effect near normal incidence across smooth potential barriers. However, the complete angular dependence of the transmission probability of massless Dirac electrons, which would be a direct signature of Klein tunneling, still remains to be measured.

Transmission resonances across SGB should therefore be accessible experimentally. The most interesting aspect would be to detect normal incidence Fabry-P\'erot resonances at zero energy and to check their strong dependence on the dimensionless momentum shift $\alpha L_x$ (see Fig. \ref{figure6}).

\section{Conclusion}

In conclusion, we have investigated zero-energy transmission resonances across a ballistic graphene bilayer in a  two terminal configuration as a function of a stacking default in the bilayer (either twist, slide or strain deformation). By the use of an effective two-band low-energy Hamiltonian around a single valley -- that accounts for the presence of two Dirac mini cones in the dispersion relation --, we have predicted two possible kinds of transmission resonances that occur in shifted bilayer graphene depending on the value of the stacking defect parameter $\Delta \propto \alpha^2 e^{2i \theta}$. The latter is parametrized by $\alpha$ -- corresponding to the mid-distance between the two Dirac points -- and $\theta$ -- controlling the orientation of these Dirac points in reciprocal space. Our key finding -- occuring when Dirac cones are hit in succession ($\theta=0$), at normal incidence and for quantized values of the momentum shift parameter $\alpha L_x= m\pi$, where $m$ is a positive integer -- is a new kind of transmission resonances of the Fabry-P\'erot type. They are unique to the shifted graphene bilayer and can be distinguished from that occurring in both undoped graphene monolayer and undoped $AB$ stacked bilayer. As soon as the shift parameter $\alpha$ differs from $m \pi/L_x$, Fabry-P\'erot resonances are suppressed in favor of evanescent waves resonances, akin to Klein tunneling, and taking place at incidence angles controlled by the energy shift parameter $\Delta (\alpha,\theta)$.

The present study provides a way to modulate the electron transmission through such graphene based device. A controllable transmission, as well as the detection of the perfect transmission, can be realized conveniently by tuning either the stacking structure and/or the incidence angle. The transmission probability through the undoped shifted graphene bilayer can be considered as a fingerprint of the stacking defect.

\section*{Acknowledgments}
We would like to thank G. Montambaux for important initial inputs and for comments on the manuscript. We also acknowledge useful discussions with F. Pi\'echon and M.O. Goerbig. This work was partially supported by the Tunisian-French CMCU 10G1306 project.


\begin{thebibliography}{99}

\bibitem{Novoselov}K.S. Novoselov \emph{et. al.}, Nature Phys. \textbf{2}, 177 (2006).

\bibitem{McCann}E. McCann, and V.I. Falko, Phys. Rev. Lett.
\textbf{96}, 086805 (2006).

\bibitem{Berger} C. Berger, Z. Song, X. Li, X. Wu, N. Brown, C. Naud, D.
Mayou, T. Li, J. Hass, A.N. Marchenkov, E.H. Conrad,
P.N. First, and W.A. de Heer, Science \textbf{312}, 1191 (2006).

\bibitem{Haas1} J. Hass, R. Feng, J.E. Millan-Otoya, X. Li, M. Sprinkle,
P.N. First, W.A. de Heer, E.H. Conrad, and C. Berger,
Phys. Rev. B \textbf{75}, 214109 (2007).

\bibitem{Haas2} J. Hass, F. Varchon, J.E. Mill\'an-Otoya, M. Sprinkle, N.
Sharma, W.A. de Heer, C. Berger, P.N. First, L. Magaud,
and E.H. Conrad, Phys. Rev. Lett. \textbf{100}, 125504 (2008).

\bibitem{Luican} A. Luican, G. Li, A. Reina, J. Kong, R.R. Nair, K.S.
Novoselov, A.K. Geim, and E.Y. Andrei, Phys. Rev. Lett.\textbf{106}, 126802 (2011).

\bibitem{Li} G. Li, A. Luican, J.M.B. Lopes dos Santos, A.H. Castro
Neto, A. Reina, J. Kong, and E.Y. Andrei, Nature Phys. \textbf{6}, 109 (2009).

\bibitem{Miller1} D.L. Miller, K.D. Kubista, G.M. Rutter, M. Ruan, W.A. de Heer, P.N. First, and J.A. Stroscio, Phys. Rev. B \textbf{81}, 125427 (2010).

\bibitem{Miller2} D.L. Miller, K.D. Kubista, G.M. Rutter, M. Ruan, W.A.
de Heer, M. Kindermann, P.N. First, and J.A. Stroscio,
Nature Phys. \textbf{6}, 811 (2010).

\bibitem{Reina} A. Reina, X.T. Jia, J. Ho, D. Nezich, H.B. Son, V. Bulovic, M.S. Dresselhaus, and J. Kong, Nano Lett. \textbf{9}, 30 (2009).

\bibitem{folding} Z. Ni, Y. Wang, T. Yu, Y. You, and Z. Shen, Phys. Rev.
B \textbf{77}, 235403 (2008).

\bibitem{segregation} R. Zhao, Y. Zhang, T. Gao, Y. Gao, N. Liu, L. Fu, and Z.
Liu, Nano Res. \textbf{4}, 712 (2011).

\bibitem{unzipping} L. Xie, H. Wang, C. Jin, X. Wang, L. Jiao, K. Suenaga,
and H. Dai, J. Am. Chem. Soc. \textbf{133}, 10394 (2011).

\bibitem{Ohta} T. Ohta, A. Bostwick, T. Seyller, K. Horn, and E. Rotenberg, Science \textbf{313}, 951 (2006).

\bibitem{Min} H. Min, and A.H. McDonald, Phys. Rev. B \textbf{77}, 155416 (2008).

\bibitem{Falko1}M. Mucha-Kruczynski, I.L. Aleiner, and V.I.
Fal'ko, Phys. Rev. B \textbf{84}, 041404 (2011).

\bibitem{Lopes}J.M.B. Lopes dos Santos, N.M.R. Peres, and A.H. Castro
Neto, Phys. Rev. Lett. \textbf{99}, 256802 (2007).

\bibitem{Trambly} G. Trambly de Laissardi\'ere, D. Mayou, and L. Magaud,
Nano Lett. \textbf{10}, 804 (2010)

\bibitem{de Gail1} R. de Gail, M.O. Goerbig, F. Guinea, G. Montambaux, and A.H. Castro Neto
Phys. Rev. B  \textbf{84}, 045436 (2011).

\bibitem{de Gail2} R. de Gail, M.O. Goerbig, and G. Montambaux,Phys. Rev. B \textbf{86},
045407 (2012).

\bibitem{Shallcross} S. Shallcross, S. Sharma, E. Kandelaki, and O.A. Pankratov, Phys. Rev. B
\textbf{81}, 239904 (2010).

\bibitem{Son} Young-Woo Son, Seon-Myeong Choi, Yoon Pyo Hong, Sungjong Woo, and Seung-Hoon
Jhi, Phys. Rev. B \textbf{84}, 155410 (2011).

\bibitem{Klein} M.I. Katsnelson, K.S. Novoselov, and A.K. Geim,
Nature Phys. \textbf{2}, 620 (2006)

\bibitem{Katsnelson_monolayer} M.I. Katsnelson, Eur. Phys. J. B
\textbf{51}, 157 (2006).

\bibitem{Tworzydlo} J. Tworzydlo, B. Trauzettel, M. Titov, A. Rycerz,
and C.W.J. Beenakker, Phys. Rev. Lett. \textbf{96}, 426802 (2006).

\bibitem{Katsnelson_bilayer} M.I. Katsnelson, Eur. Phys. J. B
\textbf{52}, 151 (2006).

\bibitem{Snyman} I. Snyman, and C.W.J. Beenakker, Phys. Rev.
B \textbf{75}, 045322 (2007).

\bibitem{Fuchs} P.E. Allain, and J.N. Fuchs, Eur. Phys. J. B\textbf{83}, 301 (2011).

\bibitem{Rickhaus}P. Rickhaus, R. Maurand, M.-H. Liu, K. Richter and C. Sch\"onenberger, Nat. Commun. 4:2342 (2013).

\bibitem{Vafek} O. Vafek, and K. Yang, Phys. Rev. B \textbf{81}, 041401 (2010).

\bibitem{Brown} L. Brown, R. Hovden, P. Huang, M. Wojcik, D.A. Muller, and J. Park, Nano Lett. \textbf{3}, 1609-15 (2012).

\bibitem{Young}A.F. Young and P. Kim, Nat. Phys. \textbf{5}, 222 (2009).

\bibitem{Cho}S. Cho and M.S. Fuhrer, Nano Res. \textbf{4}, 385 (2011).

\bibitem{Huard}B. Huard {\it et al.}, Phys. Rev. Lett. \textbf{98}, 236803 (2007).

\bibitem{Rahman} A. Rahman, J.W. Guikema, and N. Markovi\'c, arXiv:1304.5533.

\end{thebibliography}
\end{document}